\documentclass[a4paper,12pt]{article}

\usepackage{graphicx}

\usepackage[english]{babel}
\usepackage[latin1]{inputenc}
\usepackage[T1]{fontenc}
\usepackage{microtype, booktabs, graphicx, amsmath, amssymb, mathtools, braket, amsthm, aeguill, %natbib,
 amscd, sectsty, fancyhdr, eucal, chngpage, dsfont, bm, longtable}
\usepackage[bottom]{footmisc}
\usepackage[hang, small, bf, margin=20pt, tableposition=top, textfont=it]{caption}
\usepackage[inner=1.8in, outer=1.3in]{geometry}
\usepackage{times}
\usepackage{float, epstopdf}

\renewcommand{\epsilon}{\varepsilon}
\renewcommand{\phi}{\varphi}
\renewcommand{\theta}{\vartheta}

\chapterfont{\sc\centering\thispagestyle{empty}}
\chaptertitlefont{\centering}

\makeatletter
\def\cleardoublepage{\clearpage\if@twoside \ifodd\c@page\else
\hbox{}
\thispagestyle{empty}
\newpage
\if@twocolumn\hbox{}\newpage\fi\fi\fi}
\makeatother

\begin{document}

\title{On the principle of competitive exclusion in metapopulation models}
\author{Davide Belocchio, Roberto Cavoretto, Giacomo Gimmelli,\\
Alessandro Marchino, Ezio Venturino\\
Dipartimento di Matematica ``Giuseppe Peano'',\\
Universit\`a di Torino,\\
via Carlo Alberto 10, 10123 Torino, Italy\\
emails: davide.belocchio@alice.it, roberto.cavoretto@unito.it,\\
giacomo.gimmelli@virgilio.it,
marchino.alessandro@gmail.com,\\
ezio.venturino@unito.it
}
\date{}
\maketitle

\begin{abstract}
In this paper we present and analyse a simple two populations model with migrations 
among two different environments. The populations interact by competing for resources.
Equilibria are investigated. A proof for the boundedness of the populations is provided.
A kind of competitive exclusion principle for metapopulation systems is obtained.
At the same time we show that the competitive exclusion principle at the local patch level may be
prevented to hold by the migration phenomenon, i.e. two competing populations may coexist, provided
that only one of them is allowed to freely move or that migrations for both occur just in one direction.
\end{abstract}



\section{Introduction}

In this paper we consider a minimal metapopulation model with two competing populations. It consists of two different environments among which
migrations are allowed.

As migrations do occur indeed in nature, \cite{C03},
the metapopulation tool has been proposed to study populations living in fragmented habitats, \cite{Lev,SB}. One of its most important results is the
fact that a population can survive at the global level, while becoming locally extinct, \cite{H85,HG,H96,L97,M95,M98,W96}.
%,W97}.
An earlier, related concept, is the one of population assembly, \cite{LM}, to account for heterogeneous environments containing distinct
community compositions, providing insights into issues such as biodiversity and conservation. As a result, sequential slow
invasion and extinction
shape successive
species mixes into a persistent configuration, impenetrable by other species, \cite{LPNP},
while, with faster invasions,
communities change their compositions and each species has a chance to survive.

A specific example in nature for our competition situation for instance is
provided by {\it Strix occidentalis}, which competes with, and ofter succumbs to, the larger
great horned owl, {\it Bubo virginianus}. The two in fact compete for resources, since they
share several prey, \cite{HMWPH}. If the environment in which they live gets fragmented, the competition cannot be analysed classically,
and the metapopulation concept becomes essential to describe the natural interactions.
This paper attempts the development of such an issue in this framework.
Note that another recent contribution in the context of patchy environments
considers also a transmissible disease affecting the
populations, thereby introducing the concept of metaecoepidemic models, \cite{EV}.

An interesting competition metapopulation model with immediate patch occupancy by the strongest population and incorporating patch dynamics has been
proposed and investigated in \cite{MVM}. Patches are created and destroyed dynamically at different rates. A completely different approach is instead
taken for instance in \cite{CE}, where different competition models, including facilitation, inhibition and tolerance,
are investigated by means of cellular automata.

The model we study bears close resemblance with a former model recently appeared in the literature, \cite{NBZA}. However, there are two basic
distinctions, in the formulation and in the analysis. As for the model formulation, in \cite{NBZA} the populations are assumed to be
similar species competing for an implicit resource. Thus there is a unique carrying capacity for both of them in each patch in which they reside.
Furthermore their reproduction rates are the same.
We remove both these assumptions, by allowing in each patch different carrying capacities for each population, as well different reproduction rates.
Methodologically, the approach used in \cite{NBZA} uses the aggregation method, thereby reducing the system of four differential equations
to a two-dimensional one, by assuming that migrations occur at a different, faster, timescale than the life processes.
This may or may not be the case in real life situations. In fact,
referring to the herbivores inhabiting the African savannas, this movement occurs throughout the lifetime, while
intermingling for them
does not constitute a ``social'' problem, other than the standard intraspecific competition for the resources, \cite{Thaker10,Thaker11}.
The herbivores wander in search of new pastures, and the predators follow them. This behavior might instead
also be influenced by the presence of predators in the surrounding areas, \cite{Valeix}.
Thus the structure of African herbivores and the savanna ecosystems
may very well be in fact shaped by predators' behavior.

In the current classical literature in this context, it is commonly assumed that migrations of competing populations in a patchy environment
lead to the situation in which the superior competitor replaces the inferior one. In addition, it is allowed for an
inferior competitor to invade an empty patch, but the invasion is generally
prevented by the presence of a superior competitor in the patch, \cite{Til}.
Based on this setting, models investigating the proportions of
patches occupied by the superior and inferior competitors have been set up, \cite{HMcA}. The effect of patch removal in
this context is analysed in \cite{NM}, coexistence is considered in \cite{H83,Ama,RPL,AW},
habitat disruptions in a realisting setting are instead studied in \cite{M95}.
Note that in this context, the migrations are always assumed to be bidirectional. 
Our interest here differs a bit, since we want to consider also human artifacts or natural events that fragment the landscape, and therefore we will
examine particular subsystems in which migrations occur only in one direction, or
are forbidden for one of the species, due to some environmental constraints.

Our analysis shows two interesting results. First of all, a
kind of competitive exclusion principle for metapopulation systems also holds in suitable conditions.
Further, the competitive exclusion principle at the local patch level may be
overcome by the migration phenomenon, i.e. two competing populations may coexist, provided
that either only one of them is allowed to freely move, or that migrations for both populations occur just in one and the same direction.
This shows that the assumptions of the classical literature of patchy environments may at times not hold, and this remark might open up new lines
of investigations.

The paper is organized as follows. In the next Section we formulate the model showing the boundedness of its trajectories.
We proceed then to examine a few special cases, before studying the complete model: in Section \ref{one} only one population is allowed
to migrate, in Section \ref{direc} the migrations occur only in one direction. Then the full model is considered in the following Section.
A final discussion concludes the paper.

\section{Model formulation}

We consider two environments among which two competing populations can migrate, denoted by $P$ and $Q$.
Let $P_i$, $Q_i$, $i=1,2$, their sizes in the two environments. Here the subscripts
denote the environments in which they live.
Let each population thrive in each environment according to logistic growth, with possibly differing reproduction rates,
respectively $r_i$ for $P_i$ and $s_i$ for $Q_i$, and carrying capacities, respectively again $K_i$ for $P_i$ and $H_i$ for $Q_i$.
The latter are assumed to be different since they may indeed be influenced by the environment. Further let
$a_i$ denote the interspecific competition rate for $P_i$ due to the presence of the population $Q_i$ and
$b_i$ denote conversely the interspecific competition rate for $Q_i$ due to the presence of the population $P_i$.

Let $m_{ij}$ the migration rate from environment $j$ to environment $i$ for the population $P_j$ and similarly let
$n_{ij}$ be the migration rate from $j$ to $i$ for the population $Q_j$.

The resulting model has the following form:
\begin{eqnarray}\label{sistema}
\dot{P}_1 = r_1P_1\left(1-\frac{P_1}{K_1}\right)-a_1P_1Q_1-m_{21}P_1+m_{12}P_2\equiv F_1(P_1,P_2,Q_1,Q_2),\\ \nonumber
\dot{Q}_1 = s_1Q_1\left(1-\frac{Q_1}{H_1}\right)-b_1Q_1P_1-n_{21}Q_1+n_{12}Q_2\equiv F_2(P_1,P_2,Q_1,Q_2),\\ \nonumber
\dot{P}_2 = r_2P_2\left(1-\frac{P_2}{K_2}\right)-a_2P_2Q_2-m_{12}P_2+m_{21}P_1\equiv F_3(P_1,P_2,Q_1,Q_2),\\ \nonumber
\dot{Q}_2 = s_2Q_2\left(1-\frac{Q_2}{H_2}\right)-b_2Q_2P_2-n_{12}Q_2+n_{21}Q_1\equiv F_4(P_1,P_2,Q_1,Q_2).
\end{eqnarray}

Note that a very similar model has been presented in \cite{NBZA}. But (\ref{sistema}) is more general, in that it allows different carrying capacities
in the two patches for the two populations, while in \cite{NBZA} only one, $K$, is used, for both environments and populations. Further,
the environments do not affect the growth rates of each individual population, while here we allow different reproduction rates for the same population
in each different patch. Also, competition rates in \cite{NBZA} are the same in both patches, while here they are environment-dependent.
The analysis technique used in \cite{NBZA} also makes the assumption that there are two time scales in the model, the fast dynamics being represented
by migrations and the slow one by the demographics, reproduction and competition. Based on this assumption, the system is reduced to a planar one,
by at first calculating the equilibria of the fast part of the system using the aggregation method, and then the aggregated two-population slow part is analysed.

Here we thus remove the assumption of a fast migration, compared with the longer lifetime population dynamics
because for the large herbivores
the migration process is a lifelong task, being always in search of new pastures \cite{F,Valeix}.
In different environments the resources are obviously
different, making the statement on different carrying capacities more closely related to reality.
Finally, it is also more realistic to assume different carrying capacities for the two populations, even though they compete for resources, as in many cases
the competition is only partial, in the sense that their habitats overlap, but do not completely coincide.

We will consider several subcases of this system, and finally analyse it in its generality. Table \ref{tab:1}
defines all possible equilibria of the system (\ref{sistema}) together with the indication of the models
in which they appear.
For each different model examined in what follows,
we will implicitly refer to it frequently,
with only changes of notation and possibly of population levels, but not for the structure of the equilibrium,
i.e. the presence and absence of each individual population.

\begin{table}
\begin{center}
\begin{tabular}{|l|cccc||c|c|c|}
\hline
 & & & & & one migrating & migrations & full \\
 & $P_1$ & $Q_1$ & $P_2$ & $Q_2$ & population ($P$)&  $1 \rightarrow 2$ & model\\
  \hline
 $E_1$     &  0  & 0 & 0 & 0 & Y (unstable) & Y (unstable) & Y (unstable) \\
 $E_2$     &  0  & 0 & + & 0 & -- & Y & --\\
 $E_3$     &  0  & 0 & 0 & + & Y (unstable) & Y & --\\
 $E_4$     &  0  & 0 & + & + & -- & Y & --\\
 $E_5$     &  +  & 0 & 0 & 0 & -- & -- & --\\
 $E_6$     &  +  & 0 & + & 0 & Y & Y & Y\\
 $E_7$     &  +  & 0 & 0 & + & -- & -- & --\\
 $E_8$     &  +  & 0 & + & + & Y & Y & --\\
 $E_9$     &  0  & + & 0 & 0 & Y (unstable) & -- & --\\
 $E_{10}$    &  0  & + & + & 0 & -- & -- & --\\
 $E_{11}$    &  0  & + & 0 & + & Y & Y & Y\\
 $E_{12}$    &  0  & + & + & + & -- & Y & --\\
 $E_{13}$    &  +  & + & 0 & 0 & -- & -- & --\\
 $E_{14}$    &  +  & + & + & 0 & Y & -- & --\\
 $E_{15}$    &  +  & + & 0 & + & -- & -- & --\\
 $E_{16}$    &  +  & + & + & + & Y (unstable *) & Y (critical) & Y (critical) \\
 \hline
 \end{tabular}
\end{center}
\caption{All the possible equilibria of the three ecosystems: Y means that the equilibrium is possible.
We indicated also the unconditional instability, and with a star the instability verified just numerically.
Critical means that stability is achieved only under very restrictive parameter conditions, i.e. in general
the corresponding point must be considered unstable.}
\label{tab:1}
\end{table}

For the stability analyses we will need the Jacobian of (\ref{sistema}),
\begin{equation}\label{Jac}
\begin{pmatrix}
J_{11} & -a_1{P}_1 & m_{12} & 0\\
-b_1{Q}_1& J_{22} &0&n_{12}\\
m_{21}&0& J_{33} &-a_2{P}_2\\
0&n_{21}&-b_2{Q}_2& J_{44}
\end{pmatrix},
\end{equation}
where ${P}_i$ and ${Q}_i$ denote the generic equilibrium point and
\begin{align*}
&J_{11}=r_1\left(1-\frac{2{P}_1}{K_1}\right)-a_1{Q}_1-m_{21},&J_{22}=s_1\left(1-\frac{2{Q}_1}{H_1}\right)-b_1{P}_1-n_{21},\\
&J_{33}=r_2\left(1-\frac{2{P}_2}{K_2}\right)-a_2{Q}_2-m_{12},&J_{44}=s_2\left(1-\frac{2{Q}_2}{H_2}\right)-b_2{P}_2-n_{12}.
\end{align*}

\subsection{Boundedness of the trajectories}
We will now show that the solutions of (\ref{sistema}) are always bounded. We shall explain the proof of this assertion for the complete model,
but the same method can be used on each particular case, with obvious modifications.

Let us set $\phi=P_1+Q_1+P_2+Q_2$. Boundedness of $\phi$ implies boundedness for all
the populations, since they have to be non-negative. Adding up the system equations, we obtain a differential equation for $\phi$,
the right hand side of which can be bounded from above as follows
\begin{equation}
\label{boundedness}
\begin{split}
\dot{\phi}&=
r_1P_1\left(1-\frac{P_1}{K_1}\right)-a_1P_1Q_1+ s_1Q_1\left(1-\frac{Q_1}{H_1}\right)-b_1Q_1P_1\\
&\phantom{=}+r_2P_2\left(1-\frac{P_2}{K_2}\right)-a_2P_2Q_2 +s_2Q_2\left(1-\frac{Q_2}{H_2}\right)-b_2P_2Q_2\\
&\leq r_1P_1\left(1-\frac{P_1}{K_1}\right)+ s_1Q_1\left(1-\frac{Q_1}{H_1}\right)\\
&\phantom{=} +r_2P_2\left(1-\frac{P_2}{K_2}\right) +s_2Q_2\left(1-\frac{Q_2}{H_2}\right)\\
&=r_1P_1-\frac{r_1}{K_1}P_1^2+s_1Q_1-\frac{s_1}{H_1}Q_1^2+r_2P_2-\frac{r_2}{K_2}P_2^2+s_2Q_2-\frac{s_2}{H_2}Q_2^2.
\end{split}
\end{equation}
Let
\begin{eqnarray*}
\nu=\max\set{r_1,s_1,r_2,s_2},\\
\mu_1=\frac{\nu K_1}{r_1},\quad \mu_2=\frac{\nu H_1}{s_1},\quad \mu_3=\frac{\nu K_2}{r_2},\quad \mu_4=\frac{\nu H_2}{s_2}.
\end{eqnarray*}

Substituting in \eqref{boundedness} we find
\begin{align*}
\dot{\phi}&\leq \nu P_1-\nu\frac{P_1^2}{\mu_1}+\nu Q_1-\nu\frac{Q_1^2}{\mu_2}+\nu P_2-\nu\frac{P_2^2}{\mu_3}+\nu Q_2-\nu\frac{Q_2^2}{\mu_4}\\
&=\nu\left(P_1+Q_1+P_2+Q_2-\frac{P_1^2}{\mu_1}-\frac{Q_1^2}{\mu_2}-\frac{P_2^2}{\mu_3}-\frac{Q_2^2}{\mu_4}\right).
\end{align*}
If we set
\begin{equation*}
\mu_-=\min_i\set{\mu_i},\quad \mu_+=\max_+\set{\mu_i},\quad \tau=\frac{\mu_-^3}{4\mu_+^4}
\end{equation*}
we find
\begin{align*}
\dot{\phi}&\leq \nu\left(P_1+Q_1+P_2+Q_2- \frac{\mu_-^3}{\mu_+^4}\left(P_1^2+Q_1^2+P_2^2+Q_2^2\right) \right)\\
&\leq \nu\left(P_1+Q_1+P_2+Q_2- \frac{\mu_-^3}{4\mu_+^4}\left(P_1+Q_1+P_2+Q_2\right)^2 \right)\\
&=\nu\phi\left(1-\frac{\phi}{\tau}\right).
\end{align*}

Let us now set $P_1(0)+Q_1(0)+P_2(0)+Q_2(0)=\phi(0)=u_0$ and let $u$ be the solution of the Cauchy problem
$$
\dot{u}(t)=\nu u(t)\left(1-\frac{u(t)}{\tau}\right), \quad
u(0)=u_0.
$$

By means of the generalized Gr\"onwall inequality we have that $\phi(t)\leq u(t)$ for all $t>0$, and so
\begin{equation*}
\limsup_{t\to+\infty}\phi(t)\leq\limsup_{t\to+\infty} u(t)=\tau<+\infty.
\end{equation*}
This implies at once that $\phi$ is bounded, and thus the boundedness of the system's populations as desired.

Observe that the boundedness result obtained here for this minimal model is easily
generalized to meta-populations living in $n$ patches.

\section{One population unable to migrate}\label{one}
Here we assume that the $Q$ population cannot migrate between the two environments. This may be due to the fact that it is weaker, or that
there are natural obstacles that prevent it from reaching the other environment, while these obstacles instead can be overcome by the population $P$.
Thus each subpopulation $Q_1$
and $Q_2$ is segregated in its own patch. This assumption corresponds therefore to setting $n_{ij}=0$ into (\ref{sistema}).
In this case we will denote the system's equilibria by $\widehat E_k$, with $k=1,\ldots,16$.
It is easy to show that equilibria $\widehat E_2$, $\widehat E_4$, $\widehat E_{10}$, $\widehat E_{12}$ do not satisfy the first equilibrium
equation, and $\widehat E_5$, $\widehat E_7$, $\widehat E_{13}$, $\widehat E_{15}$ do not satisfy the third one, so that all these points are excluded
from our analysis since they are unfeasible.

At the origin, $\widehat E_1$, the Jacobian (\ref{Jac}) has the eigenvalues
\begin{eqnarray*}
\lambda_\pm=\frac{1}{2}(m_{12}+m_{21}-r_1-r_2)
				  \pm\frac{\sqrt{\Delta}}{2},\\
				  \Delta=(m_{12}+m_{21}-r_1-r_2)^2-4(r_1r_2-m_{12}r_1-m_{21}r_2)
\end{eqnarray*}
and $s_1>0$, $s_2>0$, from which its instability follows.

The point $\widehat E_3=(0,0,0,H_2)$ is unconditionally feasible, but the eigevalues of (\ref{Jac}) evaluated at $\widehat E_3$ turn out to be
\begin{eqnarray*}
\hspace{-30pt}\lambda_\pm=\frac{1}{2}(-a_2H_2-m_{12}-m_{21}+r_1+r_2)\pm\frac{\sqrt{\Delta}}{2},\\
\Delta=(a_2H_2+m_{12}+m_{21}-r_1-r_2)^2\\
-4(a_2H_2m_{21}-a_2H_2r_1-m_{12}r_1-m_{21}r_2+r_1r_2)
\end{eqnarray*}
together with $-s_2<0$, $s_1>0$, so that also $\widehat E_3$ is inconditionally unstable.

The point $\widehat E_{11}=(0,H_1,0,H_2)$ is always feasible. Two eigenvalues for (\ref{Jac}) are easily found, $-s_1<0$, $-s_2<0$. The other ones come from a
quadratic equation, for which the Routh-Hurwitz conditions reduce to
\begin{eqnarray}\label{hatE11_stab}
r_1r_2<a_1H_1+a_2H_2+ m_{12}+m_{21},\\ \nonumber
r_1r_2+m_{12} a_1H_1+m_{21} a_2H_2+ a_1 a_2H_1H_2>r_1(m_{12}+ a_2H_2) + r_2(m_{21}+ a_1H_1).
\end{eqnarray}
For parameter values satisfying these conditions then, $\widehat E_{11}$ is stable.

Equilibrium $\widehat E_9=(0,H_1,0,0)$ is always feasible, and the Jacobian (\ref{Jac}) has eigenvalues
\begin{eqnarray*}
\lambda_\pm=\frac{1}{2}(-a_1H_1-m_{12}-m_{21}+r_1+r_2)\pm\frac{\sqrt{\Delta}}{2},\\
\Delta=(a_1H_1+m_{12}+m_{21}-r_1-r_2)^2\\
-4(a_1H_1m_{12}-a_1H_1r_2-m_{12}r_1-m_{21}r_2+r_1r_2)
%\\
%\lambda_\pm=&\frac{1}{2}(-a_1H_1-m_{12}-m_{21}+r_1+r_2)\\
%&\pm\frac{\sqrt{(a_1H_1+m_{12}+m_{21}-r_1-r_2)^2-4(a_1H_1m_{12}-a_1H_1r_2-m_{12}r_1-m_{21}r_2+r_1r_2)}}{2}.
\end{eqnarray*}
again with $-s_1<0$, $s_2>0$ so that
in view of the positivity of the last eigenvalue, $\widehat E_9$ is always unstable.

Existence for the equilibrium $\widehat E_6$ can be established as an intersection of curves in the $P_1-P_2$ phase plane. The equations that define them
describe the following two convex parabolae
\begin{eqnarray*}
\Pi_1: \quad P_2(P_1)\equiv \frac 1{m_{12}}\left[ r_1P_1(1-\frac{P_1}{K_1})-m_{21}P_1\right],\\
\Pi_2: \quad P_1(P_2)\equiv \frac 1{m_{21}}\left[ r_2P_2(1-\frac{P_2}{K_2})-m_{12}P_2 \right].
\end{eqnarray*}
Both cross the coordinate axes at the origin and at another point, namely
$$
X\equiv \left(\frac 1{r_1}(r_1-m_{21}) K_1, 0\right), \quad W\equiv \left(0, \frac 1{r_2}(r_2-m_{12}) K_2\right)
$$
respectively for $\Pi_1$ and for $\Pi_2$. Now by drawing these curves it is easily seen that they always intersect in the first quadrant, independently of
the position of these points, except when both have negative coordinates. The latter case need to be scrutinized more closely.
To ensure a feasible intersection, we need to look at the parabolae
slopes at the origin. Thus, the feasible intersection exists if $\Pi_1'(0) [\Pi_2'(0)]^{-1}<1$ or, explicitly when
\begin{equation}\label{exist_parabolae}
m_{12} m_{21}> (m_{21}-r_1) (m_{12}-r_2).
\end{equation}
However, coupling this condition with the negativity of the coordinates of the above points $X$ and $W$,
intersections of the parabolae with the axes,
the condition for the feasibility of $\widehat E_6$ becomes simply
\begin{equation*}
r_1<m_{21},\quad r_2<m_{12},
\end{equation*}
which is exactly the assumption that the coordinates of the points $X$ and $W$ be negative. Hence it is automatically satisfied. Further,
in the particular case in which one or both such points coalesce into the origin, i.e. for either $r_1=m_{21}$ or $r_2=m_{12}$, is it easily seen
that the corresponding parabola is tangent to the origin and a feasible $\widehat E_6$ always exists.
In conclusion, the equilibrium $\widehat E_6$ is always feasible.

By using the Routh-Hurwitz criterion we can implicitly obtain the stability conditions as
\begin{eqnarray*}
s_2<b_2P_2,\quad s_1<b_1P_1,\\
r_1\left(1-\frac{2}{K_1}P_1\right)+r_2\left(1-\frac{2}{K_2}P_2 \right)<m_{12}+m_{21},\\
\left[r_1\left(1-\frac{2}{K_1}P_1\right)-m_{21}\right] \left[r_2\left(1-\frac{2}{K_2}P_2\right)-m_{12}\right] >m_{12}m_{21}.
\end{eqnarray*}

Numerical simulations reveal that the stability conditions are a nonempty set, we obtain
$\widehat E_{6}=(119.6503, 0, 167.4318, 0)$ for the parameter values
$r_1= 90.5792$, $r_2= 97.0593$, $s_1= 3.5712$, $s_2= 3.1833$, $K_1= 119.0779$, $K_2= 167.9703$, $H_1= 112.7548$,
$H_2= 212.7141$, $a_1= 41.5414$, $a_2= 2.6975$, $b_1= 39.7142$, $b_2= 4.1911$, $m_{12}= 0.9619$, $m_{21}= 0.9106$,
$n_{12}= 0$, $n_{21}= 0$.
%Figure \ref{fig:widehatE6}.

For the equilibrium point $\widehat E_8$ 
we can define two parabolae in the $P_1-Q_2$ plane by solving the equilibrium equation for $P_2$:
\begin{eqnarray}
\widehat \Pi_1:\quad Q_2(P_1)\equiv \frac{H_2 b_2}{m_{12} s_2} P_1\left(r_1-m_{21}-\frac{r_1}{K_1}P_1\right)+H_2,\\
\widehat \Pi_2:\quad P_1(Q_2)\equiv\frac{s_2}{b_2^2 m_{21}K_2H_2^2} \left[ (r_2s_2-a_2b_2H_2K_2)Q_2^2\right.\\
\nonumber\left.+(r_2b_2H_2K_2-2r_2s_2H_2+a_2b_2H_2^2K_2-m_{12}b_2K_2H_2)Q_2\right.\\
\nonumber\left.+(r_2s_2H_2^2-r_2b_2H_2^2K_2+m_{12}b_2H_2^2K_2)\right].
\end{eqnarray}

The first parabola intersects the $Q_2$ axis at the point $(0,H_2)$, it always has two real roots,
one of which is positive and the other negative, and has the vertex with abscissa $V=\frac 12 K_1(R_1-m_{21})r_1^{-1}$.
The second parabola intersects the $Q_2$ axis at the points
\begin{equation*}
R_1\equiv \left(0,\frac{b_2H_2K_2r_2-H_2r_2s_2+b_2H_2K_2m_{12}}{a_2,b_2H_2K_2-r_2s_2}\right),\quad R_2\equiv \left(0,H_2\right).
\end{equation*}

Given that the two parabolae always have one intersection on the boundary of the first quadrant,
we can formulate a certain number of conditions ensuring their intersection in the interior of the first quadrant.
These conditions arise from the abscissa of the vertex of $\widehat \Pi_1$, of the leading coefficient of $\widehat \Pi_2$ and by the
relative positions of the roots of $\widehat \Pi_2$. By denoting as mentioned by $V$ the abscissa of vertex of $\widehat \Pi_1$, by $L$ the leading
coefficient of $\widehat \Pi_2$ and by $R$ the ordinate of $R_1$, we have explicitly $8$ sets of conditions:
\begin{enumerate}
\item $V>0$, $L>0$, $R>H_2$: the feasibility condition reduces just to the intersection between $\widehat \Pi_2$ and the $P_1$ axis
being larger than the positive root of $\widehat \Pi_1$; explicitly,
\begin{eqnarray*}
r_1>m_{21}, \quad a_2 b_2 H_2 K_2<r_2 s_2,\\
K_1<\frac{r_1 s_2 \left(b_2 k_2 \left(m_{12}-r_2\right)+r_2 s_2\right)^2}{b_2^2 K_2 m_{21}
\left(b_2 K_2 \left(m_{12} r_1+\left(m_{21}-r_1\right) r_2\right)+\left(-m_{21}+r_1\right) r_2 s_2\right)},
\end{eqnarray*}
together with either $m_{12}\geq r_2$ or
$$
a_2 H_2+m_{12}>r_2>m_{12},\quad	K_2<\frac{r_2 s_2}{b_2 \left(-m_{12}+r_2\right)}.
$$
\item $V>0,L>0,R<H_2$: the feasibity condition is that the slope of $\widehat \Pi_2$ at the point $(0,H_2)$ be
smaller than that of $\widehat \Pi_1$ at the same point. But the value of the population $P_2$ in this case would be negative, thus this condition is unfeasible;
\item $V>0,L<0,R>H_2$: the feasibity condition requires the slope of $\widehat \Pi_2$ at the point $(0,H_2)$ to
be smaller than that of $\widehat \Pi_1$ at the same point. But the value of the population $P_2$ would then be negative, so that this condition is unfeasible;
\item $V>0,L<0,R<H_2$: in general there is no intersection point;
\item $V<0,L>0,R>H_2$: the feasibity condition states that the slope of $\widehat \Pi_2$ at the point $(0,H_2)$ be smaller
than that of $\widehat \Pi_1$ at the same point; explicitly
\begin{eqnarray*}
m_{21}>r_1,\quad r_2 s_2>a_2 b_2 H_2 K_2 \\
a_2 H_2<\frac{m_{12} r_1}{m_{21}-r_1}+r_2,\quad  a_2 H_2+m_{12}>r_2.
\end{eqnarray*}
\item $V<0,L>0,R<H_2$: for feasibility, the intersection between $\widehat \Pi_2$ and the $P_1$ axis must be larger than the positive root of $\widehat \Pi_1$;
in other words
\begin{eqnarray*}
m_{21}>r_1, \quad r_2>m_{12},\quad a_2 h_2+m_{12}<r_2,\\
K_1<\frac{r_1 s_2 \left(b_2 K_2 \left(m_{12}-r_2\right)+r_2 s_2\right)^2}{b_2^2 K_2 m_{21}
\left(b_2 K_2 \left(m_{12} r_1+\left(m_{21}-r_1\right) r_2\right)+\left(-m_{21}+r_1\right) r_2 s_2\right)},\\
\frac{\left(m_{21}-r_1\right) r_2 s_2}{b_2 \left(m_{12} r_1+\left(m_{21}-r_1\right) r_2\right)}<K_2<\frac{r_2 s_2}{b_2 \left(-m_{12}+r_2\right)}.
\end{eqnarray*}
\item $V<0,L<0,R>H_2$: there can be no intersection point;
\item $V<0,L<0,R<H_2$: for feasibity the slope of $\widehat \Pi_2$ at the point $(0,H_2)$ must be smaller than that of $\widehat \Pi_1$ at the same point.
In this case, explicitly we have the feasibility conditions
\begin{equation*}
m_{21}>r_1,\quad r_2>a_2 H_2+m_{12},\quad a_2 b_2 H_2 K_2>r_2 s_2.
\end{equation*}

\end{enumerate}
The stability conditions given by the Routh-Hurwitz criterion can be stated as $s_1<b_1P_1$ together with
\begin{eqnarray*}
%s_1<b_1P_1,\\
m_{12}+m_{21}+\frac{2 P_1 r_1}{K_1}+P_2 \left(b_2+\frac{2 r_2}{K_2}\right)+Q_2 \left(a_2+\frac{2 s_2}{H_2}\right)>r_1+r_2+s_2,\\
b_2 H_2 P_2 \left( \left(K_2-2 P_2\right) \left(K_1 \left(r_1-m_{21}\right)-2 P_1 r_1\right) r_2-K_2 m_{12} \left(K_1-2 P_1\right) r_1\right)\\
>\left(H_2-2 Q_2\right) \left(K_2 \left(a_2 K_1 m_{21} Q_2-\left(K_1-2 P_1\right) \left(m_{12}+a_2 Q_2\right) r_1\right)  \right.\\
\left.   -\left(K_2-2 P_2\right) \left(K_1 \left(m_{21}-r_1\right)+2 P_1 r_1\right) r_2\right) s_2
\end{eqnarray*}
and finally
\begin{eqnarray*}
\left(H_2 \left(2 K_2 P_1 r_1+K_1 \left(2 P_2 r_2+K_2 \left(m_{12}+m_{21}+b_2 P_2+a_2 Q_2-r_1-r_2-s_2\right)  \right)\right)\right.\\
\left.+2 K_1 K_2 Q_2 s_2\right) \left(H_2 \left(K_2 \left(a_2 K_1 m_{21} Q_2-\left(K_1-2 P_1\right) \left(m_{12}+a_2 Q_2\right) r_1\right)  \right.\right.\\
\left.-\left(K_2-2 P_2\right) \left(K_1 \left(m_{21}-r_1\right)+2 P_1 r_1\right) r_2\right)+b_2 H_2 P_2 \left(2 K_2 P_1 r_1+K_1\right.\\
\times \left.\left(K_2 \left(m_{12}+m_{21}-r_1-r_2\right) +2 P_2 r_2\right)\right)-\left(H_2-2 Q_2\right) \left(2 K_2 P_1 r_1+K_1\right.\\
\times \left.\left.\left(K_2 \left(m_{12}+m_{21}+a_2 Q_2-r_1-r_2\right)+2 P_2 r_2\right)\right) s_2\right)\\
>H_2 K_1 K_2\left(b_2 H_2 P_2 \left(-K_2 m_{12} \left(K_1-2 P_1\right)r_1\right.\right.\\
\left. +\left(K_2-2 P_2\right) \left(-2 P_1 r_1+K_1 \left(-m_{21}+r_1\right)\right) r_2\right)\\
-\left(H_2-2 Q_2\right) \left(K_2 \left(a_2 K_1 m_{21} Q_2-\left(K_1-2 P_1\right) \left(m_{12}+a_2 Q_2\right) r_1\right)\right.\\
\left.\left.-\left(K_2-2 P_2\right) \left(K_1 \left(m_{21}-r_1\right)+2 P_1 r_1\right) r_2\right) s_2\right),
\end{eqnarray*}
where the population values are those at equilibrium. Also in this case the simulations show that this equilibrium
$\widehat E_{8}=(220.0633, 0, 0.0176, 247.9334)$ can be achieved for the parameter values
$r_1= 148.9386$, $r_2= 97.3583$, $s_1= 162.3161$, $s_2= 94.1847$, $K_1= 221.5104$, $K_2= 260.2843$,
$H_1= 240.0507$, $H_2= 252.1136$, $a_1= 91.3287$, $a_2= 49.4174$, $b_1= 50.0022$, $b_2= 88.6512$,
$m_{12}= 0.0424$, $m_{21}= 0.9730$, $n_{12}= 0$, $n_{21}= 0$.

%, Figure \ref{fig:widehatE8}.

For the equilibrium $\widehat E_{14}$ the same above analysis can be repeated,
with only changes in the parabolae and in the subscripts of the above explicit feasibility conditions.
The details are omitted, but the results provide a set of feasibility conditions
\begin{equation*}
m_{12}>r_2,\quad r_1>a_1 H_1+m_{21},\quad a_1 b_1 H_1 K_1>r_1 s_1,
\end{equation*}
and the following stability conditions given by the Routh-Hurwitz criterion $s_2<b_2P_2$ together with
\begin{eqnarray*}
m_{12}+m_{21}+P_1 \left(b_1+\frac{2 r_1}{K_1}\right)+\frac{2 P_2 r_2}{K_2}+Q_1 \left(a_1+\frac{2 s_1}{H_1}\right)>r_1+r_2+s_1,\\
b_1 H_1 P_1 \left(\left(K_2-2 P_2\right) \left(K_1 \left(r_1-m_{21}\right)-2 P_1 r_1\right) r_2-K_2 m_{12} \left(K_1-2 P_1\right) r_1\right)\\
>\left(H_1-2 Q_1\right) \left(K_2 m_{12} \left(a_1 K_1 Q_1-\left(K_1-2 P_1\right) r_1\right)\right.\\
\left.-\left(K_2-2 P_2\right) \left(K_1 \left(m_{21}+a_1 Q_1-r_1\right)+2 P_1 r_1\right) r_2\right) s_1,\\
\end{eqnarray*}
and finally
\begin{eqnarray*}
\left(H_1 \left(2 K_2 P_1 r_1+K_1 \left(2 P_2 r_2+K_2 \left(m_{12}+m_{21}+b_1 P_1+a_1 Q_1-r_1-r_2-s_1\right)\right)\right)\right.\\
\left.+2 K_1 K_2 Q_1 s_1\right) \left(H_1 \left(K_2 m_{12} \left(a_1 K_1 Q_1-\left(K_1-2 P_1\right) r_1\right)-\left(K_2-2 P_2\right)\right.\right.\\
\times\left.\left.  \left(K_1 \left(m_{21}+a_1 Q_1-r_1\right)+2 P_1 r_1\right) r_2\right)+b_1 H_1 P_1\right.\\
\left. \times\left(2 K_2 P_1 r_1+K_1 \left(K_2 \left(m_{12}+m_{21}-r_1-r_2\right)+2 P_2 r_2\right)\right)-\left(H_1-2 Q_1\right)\right.\\
\times\left. \left(2 K_2 P_1 r_1+K_1 \left(K_2 \left(m_{12}+m_{21}+a_1 Q_1-r_1-r_2\right)+2 P_2 r_2\right)\right) s_1\right)\\
>H_1 K_1 K_2\left(b_1 H_1 P_1 \left(-K_2 m_{12} \left(K_1-2 P_1\right)r_1\right.\right.\\
\left.\left. +\left(K_2-2 P_2\right) \left(-2 P_1 r_1+K_1 \left(-m_{21}+r_1\right)\right) r_2\right)\right.\\
\left. -\left(H_1-2 Q_1\right) \left(K_2 m_{12} \left(a_1 K_1 Q_1-\left(K_1-2 P_1\right) r_1\right) \right. \right.\\
\left.\left. -\left(K_2-2 P_2\right)
%\right.\right.
\left(K_1 \left(m_{21}+a_1 Q_1-r_1\right)+2 P_1 r_1\right) r_2\right) s_1\right),
\end{eqnarray*}
with population values evaluated at equilibrium.
Again, the whole set of conditions can be satisfied to lead to a stable configuration for the following parameter choice:
$r_1= 19.5081$, $r_2= 28.3773$, $s_1= 151.5480$, $s_2= 164.6916$, $K_1= 224.4882$, $K_2= 249.8364$,
$H_1= 247.9646$, $H_2= 234.9984$, $a_1= 28.5839$, $a_2= 12.9906$, $b_1= 60.1982$, $b_2= 82.5817$,
$m_{12}= 0.8687$, $m_{21}= 0.1361$, $n_{12}= 0$, $n_{21}= 0$, with initial conditions $(7.5967, 48.9253, 13.1973, 16.8990)$.
The equilibrium coordinates are  $E_{14}=(0.0301, 244.9973, 242.1885, 0)$.
%, Figure~\ref{fig:widehatE14}.

The coexistence equilibrium $\widehat E_{16}$ has been deeply investigated numerically.
It has been found to be always feasible, but never stable for all the sets of parameters used.

\section{Unidirectional migration only}\label{direc}

In this case, we assume that it is not possible to migrate from patch 2 back into patch 1, so that the coefficients $m_{12}$ and $n_{12}$ vanish.
The reasons behind this statement can be found in natural situations.
For instance it can be observed that freshwater fishes swim downstream much more easily than upstream. In particular
obstacles like dams and waterfalls may hinder the upstream migrations. In any case the
overcoming of these obstacles requires a sizeable effort, for which sufficient energy must be allocated. This however may not always be available.

We denote the equilibria here by $\widetilde E_k$, $k=1,\ldots, 16$.
Equilibria $\widetilde E_5$, $\widetilde E_7$, $\widetilde E_9$, $\widetilde E_{10}$, $\widetilde E_{13}$, $\widetilde E_{14}$,
$\widetilde E_{15}$ are found to be all infeasible.

The origin $\widetilde E_1$ has two positive eigevalues $r_2>0$ and $s_2>0$, so that it is unstable.

The points $\widetilde E_2=(0,0,K_2,0)$ and $\widetilde E_3=(0,0,0,H_2)$ are feasible. For the former, the eigenvalues of the Jacobian are
$-r_2$, $-m_{21}+r_1$, $-n_{21}+s_1$, $-b_2 K_2+s_2$, giving the stability conditions
\begin{equation}\label{tildeE2_stab}
r_1<m_{21},\quad s_1<n_{21},\quad s_2<b_2K_2.
\end{equation}
For the latter instead, the eigenvalues are
$-m_{21}+r_1$, $-a_2H_2+r_2$, $-n_{21}+s_1$, $-s_2$, with the following conditional stability conditions
\begin{equation}\label{tildeE3_stab}
r_1<m_{21},\quad r_2<a_2H_2,\quad s_1<n_{21}.
\end{equation}

Equilibrium 
$$
\widetilde E_4=\left( 0,0,\frac{K_2s_2(H_2a_2-r_2)}{a_2b_2H_2K_2-r_2s_2},\frac{H_2r_2(b_2K_2-s_2)}{a_2b_2H_2K_2-r_2s_2}\right)
$$
is feasible for either one of the two alternative sets of inequalities
\begin{eqnarray}\label{tildeE4_feas_a}
a_2H_2>r_2,\quad b_2K_2>s_2; \\ \label{tildeE4_feas_b}
a_2H_2<r_2,\quad b_2K_2<s_2.
\end{eqnarray}
The eigenvalues are
$-m_{21}+r_1$, $-n_{21}+s_1$, $\lambda_{\pm}$, where
\begin{align*}
2\left(a_2b_2H_2K_2-r_2s_2\right)\lambda_\pm = r_2^2s_2+r_2s_2\left(-a_2H_2-b_2K_2+s_2\right) \pm \sqrt{\Delta},\\
\Delta=r_2s_2\left[r_2s_2\left(-a_2H_2-b_2K_2+r_2+s_2\right)2 \right.\\
\left.+4\left(a_2H_2-r_2\right)\left(b_2K_2-s_2\right)\left(a_2b_2H_2K_2-r_2s_2\right)\right]
%2( a_2 b_2 h_2 k_2- r_2 s_2)\lambda_\pm=r_2^2 s_2+r_2 s_2 \left(-a_2 h_2-b_2 k_2+s_2\right)\pm \sqrt{\Delta},\\
%\Delta=r_2 s_2 \left[r_2 s_2 \left(-a_2 h_2-b_2 k_2+r_2+s_2\right)^2 \right.\\
%\left. +4 \left(a_2 h_2-r_2\right) \left(b_2 k_2-s_2\right) \left(a_2 b_2 h_2 k_2-r_2 s_2\right)\right]
\end{align*}
In case (\ref{tildeE4_feas_a}) holds, we find $\lambda_+>0$ so that $\widetilde E_4$ is unstable. In case instead of (\ref{tildeE4_feas_b}) the
stability conditions are
\begin{equation}\label{tildeE4_stab}
r_1<m_{21},\quad s_1<n_{21},
\end{equation}
and simulations show that this point is indeed stably achieved
%, Figure \ref{fig:tildeE4}.
for the parameter values
$r_1= 0.15$, $r_2= 90$, $s_1= 0.55$, $s_2= 61$, $K_1= 250$, $K_2= 300$, $H_1= 120$, $H_2= 500$, $a_1= 12$, $a_2= 0.06$, $b_1= 3$, $b_2= 0.015$, $m_{12}= 0$, $m_{21}= 0.9$, $n_{12}= 0$, $n_{21}= 0.8$,
giving the equilibrium $\widetilde E_4=(0, 0, 205,2799, 474,9398)$.

The next points come in pairs. They are
\begin{eqnarray*}
\widetilde E_{6\pm}=
\left(\frac{K_1(r_1-m_{21})}{r_1},0,\frac{K_2r_1r_2\pm\sqrt{A}}{2r_1r_2},0\right),\\
\widetilde E_{11\pm}=
\left(0,\frac{H_1(s_1-n_{21})}{s_1},0,\frac{H_2s_1s_2\pm\sqrt{B}}{2s_1s_2}\right),
\end{eqnarray*}
where
\begin{eqnarray*}
A=&K_2r_1r_2(-4K_1m_{21}^2+4K_1m_{21}r_1+K_2r_1r_2),\\
B=&H_2s_1s_2(-4H_1n_{21}^2+4H_1n_{21}s_1+H_2s_1s_2),
\end{eqnarray*}
and with respective conditions for the non-negativity of their first components given by
\begin{eqnarray}\label{tildeE6_feas}
r_1\ge m_{21},\\ \label{tildeE11_feas}
s_1\ge n_{21}.
\end{eqnarray}
Note further that if (\ref{tildeE6_feas}) and (\ref{tildeE11_feas}) hold, then $A,B>0$.
But then $\sqrt A>K_2r_1r_2$ and $\sqrt B>H_2s_1s_2$, so that $\widetilde E_{6-}$ and $\widetilde E_{11-}$ have the second component negative, i.e. they
are infeasible. The feasibility conditions for $\widetilde E_{6+}$ and $\widetilde E_{11+}$ are then respectively
given by (\ref{tildeE6_feas}) and (\ref{tildeE11_feas}).
The eigenvalues for $\widetilde E_{6+}$ are $m_{21}-r_1$ and
\begin{equation*}
-n_{21}+\frac{b_1 K_1 \left(m_{21}-r_1\right)}{r_1}+s_1,\quad
-\frac{\sqrt{A}}{K_2 r_1},\quad \frac{b_2 \left(-K_2 r_1 r_2-\sqrt{A}\right)}{2 r_1 r_2}+s_2.
\end{equation*}
giving the stability conditions
\begin{equation}\label{tildeE6+_stab}
%r_1>m_{21},\quad 
r_1(n_{21}-s_1)>b_1K_1(m_{21}-r_1),\quad 2r_1 r_2s_2<b_2\left( K_2 r_1 r_2+\sqrt A \right),
\end{equation}
where we used (\ref{tildeE6_feas}).

Eigenvalues of $\widetilde E_{11+}$ are $n_{21}-s_1$ and
\begin{equation*}
-m_{21}+r_1+\frac{a_1 H_1 \left(n_{21}-s_1\right)}{s_1},\quad
-\frac{\sqrt{B}}{H_2 s_1},\quad r_2+\frac{1}{2} a_2 \left(-H_2-\frac{\sqrt{B}}{s_1 s_2}\right)
\end{equation*}
from which the stability conditions follow
\begin{equation}\label{tildeE11+_stab}
s_1(m_{21}-r_1)>a_1H_1(n_{21}-s_1), \quad 2r_2 s_1 s_2<a_2 \left( H_2 s_1 s_2+ \sqrt B \right).
\end{equation}
having again used (\ref{tildeE11_feas}).

For the next two equilibria, we are able only to analyse feasibility. We find
$$
\widetilde E_8=\left( \frac{-K_1 m_{21} + K_1 r_1}{r_1}, 0,B,A \right)
$$
with
\begin{align*}
A=& \frac 1{2 (a_2 b_2 H_2 K_2 r_1 - r_1 r_2 s_2)} \left\{ a_2 H_2 K_2 r_1 s_2 - 
    K_2 r_1 r_2 s_2 \right. \\
    & +\left[ -4 (K_1 K_2 m_{21}^2 s_2 - 
          K_1 K_2 m_{21} r_1 s_2) (-a_2 b_2 H_2 K_2 r_1 + 
          r_1 r_2 s_2) \right. \\
          &\left. \left.  +(a_2 H_2 K_2 r_1 s_2 - 
         K_2 r_1 r_2 s_2)^2\right]^{1/2} \right\}\\
      B=&\frac{1}{2s_2(a_2 b_2 H_2 K_2 r_1 - r_1 r_2 s_2)} \left\{ H_2 s_2 - \frac{a_2 b_2 H_2^2 K_2 r_1 s_2+b_2 H_2 K_2 r_1 r_2 s_2}{
   2 (a_2 b_2 H_2 K_2 r_1 - r_1 r_2 s_2)}\right.\\
   & - \left[ b_2 H_2 (-4 (K_1 K_2 m_{21}^2 s_2 - 
            K_1 K_2 m_{21} r_1 s_2) (-a_2 b_2 H_2 K_2 r_1 + 
            r_1 r_2 s_2) \right.\\
            & \left. \left. + (a_2 H_2 K_2 r_1 s_2 - 
           K_2 r_1 r_2 s_2)^2\right]^{1/2}\right\},
\end{align*}
and
$$
\widetilde E_{12}=\left( 0,\frac{-H_1n_{21}+H_1s_1}{s_1},D,C\right)
$$
where
\begin{align*}
D=& \frac 1 {2 (b_2 a_2 K_2 H_2 s_1 - s_1 s_2 r_2)} \left\{ b_2 K_2 H_2 s_1 r_2 - 
    H_2 s_1 s_2 r_2 \right. \\
    & +\left[ -4 (H_1 H_2 n_{21}^2 r_2 - 
          H_1 H_2 n_{21} s_1 r_2) (-b_2 a_2 K_2 H_2 s_1 + 
          s_1 s_2 r_2) \right. \\
          & \left. \left. +(b_2 K_2 H_2 s_1 r_2 - 
         H_2 s_1 s_2 r_2)^2\right]^{1/2}\right\}\\
C=&\frac{1}{2 r_2 (b_2 a_2 K_2 H_2 s_1 - s_1 s_2 r_2)} \left\{ K_2 r_2 - \frac{b_2 a_2 K_2^2 H_2 s_1 r_2+a_2 K_2 H_2 s_1 s_2 r_2}{
   2 (b_2 a_2 K_2 H_2 s_1 - s_1 s_2 r_2)} \right. \\
   & - \left[ a_2 K_2 (-4 (H_1 H_2 n_{21}^2 r_2 - 
            H_1 H_2 n_{21} s_1 r_2) (-b_2 a_2 K_2 H_2 s_1 + 
            s_1 s_2 r_2) \right. \\
            & \left. \left. + (b_2 K_2 H_2 s_1 r_2 - 
           H_2 s_1 s_2 r_2)^2\right]^{1/2}\right\}.
\end{align*}
Feasibility for $\widetilde E_8$ is ensured by
\begin{equation}\label{tildeE8_feas}
r_1>m_{21},\quad A>0,\quad B>0,
\end{equation}
while for $\widetilde E_{12}$ by
\begin{equation}\label{tildeE12_feas}
s_1>n_{21},\quad C>0,\quad D>0.
\end{equation}
Numerical simulations show in fact their stability,
% as seen in Figures \ref{fig:tildeE8} and \ref{fig:tildeE12}.
respectively for the parameter values
$r_1= 148.8149$, $r_2= 95.9844$, $s_1= 121.9733$, $s_2= 171.8885$, $K_1= 228.8361$, $K_2= 223.9932$, $H_1= 201.4337$, $H_2= 216.7927$, $a_1= 71.2694$, $a_2= 47.1088$, $b_1= 68.1972$, $b_2= 7.1445$, $m_{12}= 0$, $m_{21}= 0.8175$, $n_{12}= 0$, $n_{21}= 0.5186$, giving
%with initial conditions $(64.8991, 45.3798, 82.5314, 13.3171)$. The 
equilibrium $\widehat E_8=(227.5790, 0, 0.0184, 216.6269)$
and
for the parameter values
$r_1= 70.3319$, $r_2= 117.0528$, $s_1= 183.4387$, $s_2= 151.4400$, $K_1= 219.0223$, $K_2= 207.5854$, $H_1= 226.5399$, $H_2= 293.4011$, $a_1= 56.8824$, $a_2= 1.1902$, $b_1= 16.2182$, $b_2= 31.1215$, $m_{12}= 0$, $m_{21}= 0.2630$, $n_{12}= 0$, $n_{21}= 0.4505$, giving 
%, with initial conditions $(22.8977, 15.2378, 53.8342, 7.8176)$. The equilibrium coordinates are 
$\widetilde E_{12}=(0, 225.9835, 207.5514, 0.0161)$.

The coexistence equilibrium $\widetilde E_{16}=(18.4266, 18.4266, 18.6164, 18.6164)$
has been numerically investigated for the parameter values
$r_1= 100$, $r_2= 100$, $s_1= 100$, $s_2= 100$, $K_1= 250$, $K_2= 250$, $H_1= 250$, $H_2= 250$,
$a_1= 5$, $a_2= 5$, $b_1= 5$, $b_2= 5$, $m_{12}= 0$, $m_{21}= 0.5$, $n_{12}= 0$, $n_{21}= 0.5$.
% in Figure \ref{fig:tildeE16},
from which its stability under suitable parameter values is shown. Note that the parameters have been chosen
in a very peculiar way, the reproduction rates all coincide, as do all the carrying capacities, the competition
rates and the migration rates. However, numerical experiments reveal that by slightly perturbing these values, 
the stability of this equilibrium point is immediately lost. We conclude then that the coexistence equilibrium
can be achieved at times, but is generically unstable.

\section{The complete model}

We consider now the full system (\ref{sistema})
In this case, the points
$E_2$, $E_3$, $E_4$, $E_5$, $E_7$, $E_8$, $E_9$, $E_{10}$, $E_{12}$, $E_{13}$, $E_{14}$, $E_{15}$ are seen to be all infeasible.

At the origin $E_1$, the characteristic polynomial factors to give the two quadratic equations
$$
\lambda^2 -\lambda (r_1-m_{21}+r_2-m_{12}) + (r_1-m_{21})(r_2-m_{12})-m_{21}m_{12}=0
$$
and
$$
\lambda^2 -\lambda (s_1-n_{21}+s_2-n_{12}) + (s_1-n_{21})(s_2-n_{12})-n_{21}n_{12}=0.
$$
Stability conditions are then ensured by the Routh-Hurwitz conditions, which explicitly become
\begin{eqnarray}\label{E1_stab}
m_{21}+m_{12}>r_1+r_2, \quad r_1r_2> r_1 m_{12} + r_2 m_{21}, \\ \nonumber
n_{21}+n_{12}>s_1+s_2, \quad s_1s_2> s_1 n_{12} + s_2 n_{21}.
\end{eqnarray}

These conditions are nevertheless incompatible, since from the second one we have $r_1>m_{21}+m_{12}r_1 r_2^{-1}>m_{21}$ and similarly $r_2>m_{12}$,
contradicting thus the first one.
The origin is therefore always unstable.

The points $E_6$ and $E_{11}$ may be studied by the same means of \eqref{exist_parabolae} and therefore are always feasible. The stability of $E_6$ is given implicitly by
\begin{eqnarray*}
s_2<b_2P_2,\quad s_1<b_1P_1,\\
r_1\left(1-\frac{2}{K_1}P_1\right)+r_2\left(1-\frac{2}{K_2}P_2 \right)<m_{12}+m_{21},\\
\left[r_1\left(1-\frac{2}{K_1}P_1\right)-m_{21}\right] \left[r_2\left(1-\frac{2}{K_2}P_2\right)-m_{12}\right] >m_{12}m_{21},
\end{eqnarray*}
whereas for the equilibrium $E_{11}$ we have the conditions
\begin{eqnarray*}
r_2<a_2Q_2,\quad r_1<a_1Q_1,\\
s_1\left(1-\frac{2}{H_1}Q_1\right)+s_2\left(1-\frac{2}{H_2}Q_2 \right)<n_{12}+n_{21},\\
\left[s_1\left(1-\frac{2}{H_1}Q_1\right)-n_{21}\right] \left[s_2\left(1-\frac{2}{H_2}Q_2\right)-n_{12}\right] >n_{12}n_{21}.
\end{eqnarray*}
Simulations were carried out to demonstrate that the stability conditions of these points can be satisfied.
The equilibrium $E_6=(203.2749, 0, 262.4315, 0)$ is stably achieved for the parameter values
$r_1= 179.2820$, $r_2= 48.8346$, $s_1= 162.9841$, $s_2= 27.9518$, $K_1= 202.4929$, $K_2= 265.3457$,
$H_1= 204.9169$, $H_2= 203.4834$, $a_1= 58.2431$, $a_2= 69.7650$, $b_1= 94.4784$, $b_2= 77.2208$,
$m_{12}= 0.9758$, $m_{21}= 0.5674$, $n_{12}= 0.4716$, $n_{21}= 0.2537$.
%, with initial conditions $(85.6351, 13.2285, 52.6980, 47.9404)$. The equilibrium coordinates are  .
Equilibrium $E_{11}=(0, 245.4094, 0, 263.5643)$ is attained with the choice
$r_1= 46.2191$, $r_2= 191.5950$, $s_1= 70.5120$, $s_2= 171.4748$, $K_1= 240.3233$,
$K_2= 256.7841$, $H_1= 244.9968$, $H_2= 263.8244$, $a_1= 49.1146$, $a_2= 43.3295$,
$b_1= 77.5334$, $b_2= 38.0149$, $m_{12}= 0.4620$, $m_{21}= 0.6463$, $n_{12}= 0.8896$,
$n_{21}= 0.8370$, with initial conditions $(47.4215, 86.4803, 27.8785, 70.8909)$.
%The equilibrium coordinates are  $E_{11}=(0, 245.4094, 0, 263.5643)$.
%as can be seen in Figures \ref{fig:E6} and \ref{fig:E11}.

For the coexistence equilibrium $E_{16}=(10.7367, 10.7367, 15.0240, 15.0240)$ we have similar results as for the one of the one-migration only case.
It
% is shown to 
exists and is stable
%in Figure \ref{fig:E16}, 
for the very specific parameter values
$r_1= 110$, $r_2= 80$,
$s_1= 110$, $s_2= 80$, $K_1= 360$, $K_2= 270$, $H_1= 360$, $H_2= 270$, $a_1= 10$, $a_2= 5$, $b_1= 10$,
$b_2= 5$, $m_{12}= 0.5$, $m_{21}= 0.1$, $n_{12}= 0.5$, $n_{21}= 0.1$.
Its stability however is easily broken under slight perturbations of the 
system parameters. Again, thus, the coexistence equilibrium $E_{16}$ is not generically stable.

\section{Conclusions}

\subsection{Discussion of the possible systems' equilibria}
The metapopulation models of competition type here considered show that only a few populations configurations are possible at a stable level.
First of all, in virtue of our assumptions, all these ecosystems will never disappear.
Table \ref{tab:1} shows that equilibria $E_{5}$, $E_{7}$, $E_{10}$, $E_{13}$, $E_{15}$ cannot occur in any one of the models considered here.
Of these, $E_{7}$ and $E_{10}$ are the most interesting ones. They show that one competitor cannot survive solely in one patch, while the other one
thrives alone in the second patch. Thus it is not possible to reverse the outcome of a superior competitor in one patch in the other patch.
Further, in the first patch the two populations can coexist only in the model
in which only one population is allowed to migrate back and forth into the other patch, equilibrium $E_{14}$.
In that case, the migrating population thrives also alone in the second environment.
The coexistence of all populations in both environments is ``fragile'', it occurs only under very limited assumptions.
Coexistence in the second patch can occur instead with the first one empty at $E_{4}$, only
in the following two cases. For the one-directional migration model, with immigrations into the second patch,
the first patch is left empty.
When the first patch is instead populated by one species only, at
equilibria $E_{8}$ for both the one-population and unidirectional migrations models
and at $E_{12}$, again for the one-directional
migrations model.
The equilibria in which one population is wiped out from the ecosystem instead, $E_{6}$ and $E_{11}$, occur in all three models.
Finally, the three remaining equilibria contain only one population in just one patch.
At $E_{2}$, only for the unidirectional migration model, the migrating population survives in the arrival patch.
At $E_{9}$ it is the residential, i.e. the non-migrating,
population that survives in its own patch, only for the one-population migrations model.
At $E_{3}$ for both particular cases instead, the residential population survives in the ``arrival'' patch of the other migrating population.

\subsection{Unrestricted migrations}
Looking now more specifically at each one of the proposed models, we draw the following inferences.

The model with unrestricted migration possibilities allows the survival of either one of the competing populations, in both patches,
$E_6$ and $E_{11}$. Coupling
this result with the fact that the interior coexistence has been numerically shown to be stable
just for a specific parameter choice, but it is
generally unstable, this result appears
to be an extension of the classical competitive exclusion principle, \cite{MPV}, to metapopulation systems,
in agreement with the classical literature in the field, e.g. \cite{AW,Ama,H83,HMcA,M95,NM,RPL,Til}. It is apparent here, as well as in the classical case,
that an information on how the basins of attraction of the two mutually exclusive boundary equilibria is important in assessing the final
outcome of the system, based on the knowledge of its present state. To this end, relevant numerical work has been performed for two
dimensional systems, \cite{CCDRM}. An extension to higher dimensions is in progress.

\subsection{Migration allowed for just one population.}
For the model in which only one population can migrate, two more equilibria are possible
in addition to those of the full model, i.e. the 
resident, non-migrating, population $Q$ can survive just
in one patch with the migrating one, and the patch can be either one of the two in the model, equilibria $\widehat E_8$ and $\widehat E_{14}$. The resident population cannot outcompete the migrating one, since
the equilibria $\widehat E_3$ and $\widehat E_{9}$ are both unconditionally unstable.
Thus, when just one population migrates, the classical principle of competitive exclusion does not necessarily hold
neither at the wider metapopulation level,
nor in one of the two patches, as shown by the nonvanishing population levels of patch 2 in 
equilibrium $\widehat E_{8}=(220.0633, 0, 0.0176, 247.9334)$
%Figure \ref{fig:widehatE8} 
and in patch 1
in equilibrium $\widehat E_{14}=(0.0301, 244.9973, 242.1885, 0)$.
%Figure \ref{fig:widehatE14}.
The coexistence in one of the two patches
appears to be possible since the weaker species can migrate to the other competitor-free environment, thrive there and
migrate back to reestablish itself with the competitor in the original environment.
But the principle of competitive exclusion can in fact occur also in this model, since the numerical
simulations reveal it, consider indeed the equilibrium $\widehat E_{6}=(119.6503, 0, 167.4318, 0)$.
% found in Figure \ref{fig:widehatE6}.
However, restrictions in the interpatch moving possibilities of one
population might prevent its occurrence. The coexistence of all the populations appears to be always impossible in view of the instability
of the equilibrium $\widehat E_{16}$.

Using the algorithm introduced in \cite{CCDRM},
we have also explored a bit how the migration rates influence the shape of the basins of attraction of the two
equilibria $\widehat E_6$ and $\widehat E_{11}$.

\begin{figure}[p]
\centering
\includegraphics[width=5.0cm]{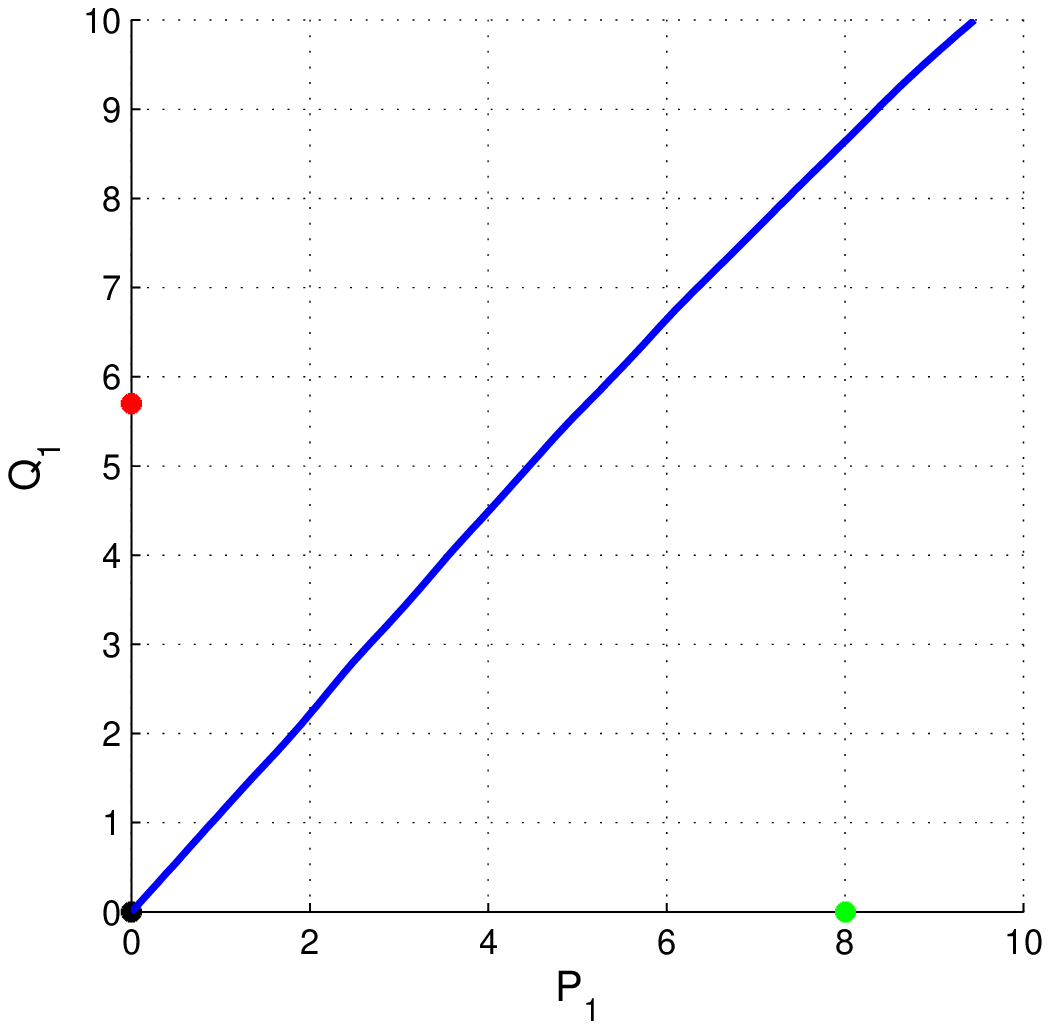}
\includegraphics[width=5.0cm]{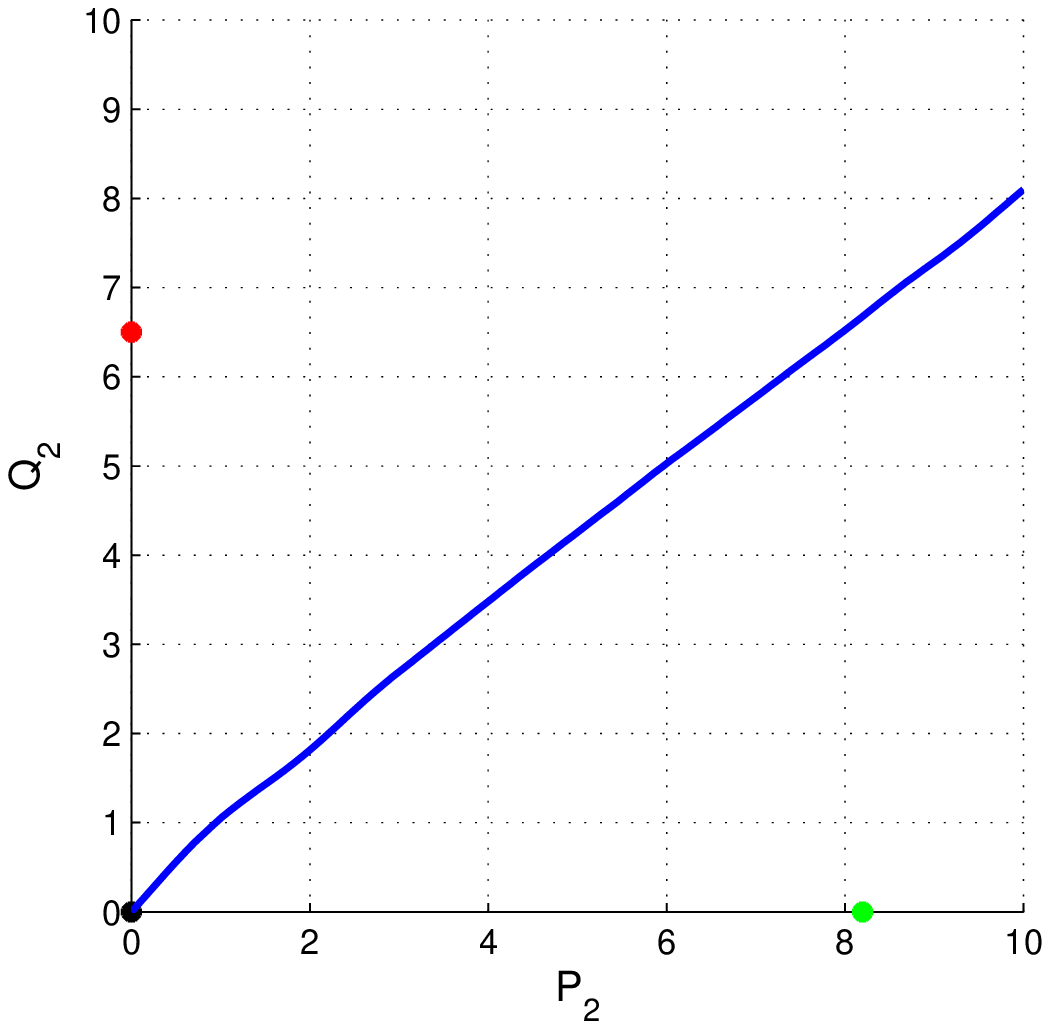}
\includegraphics[width=5.0cm]{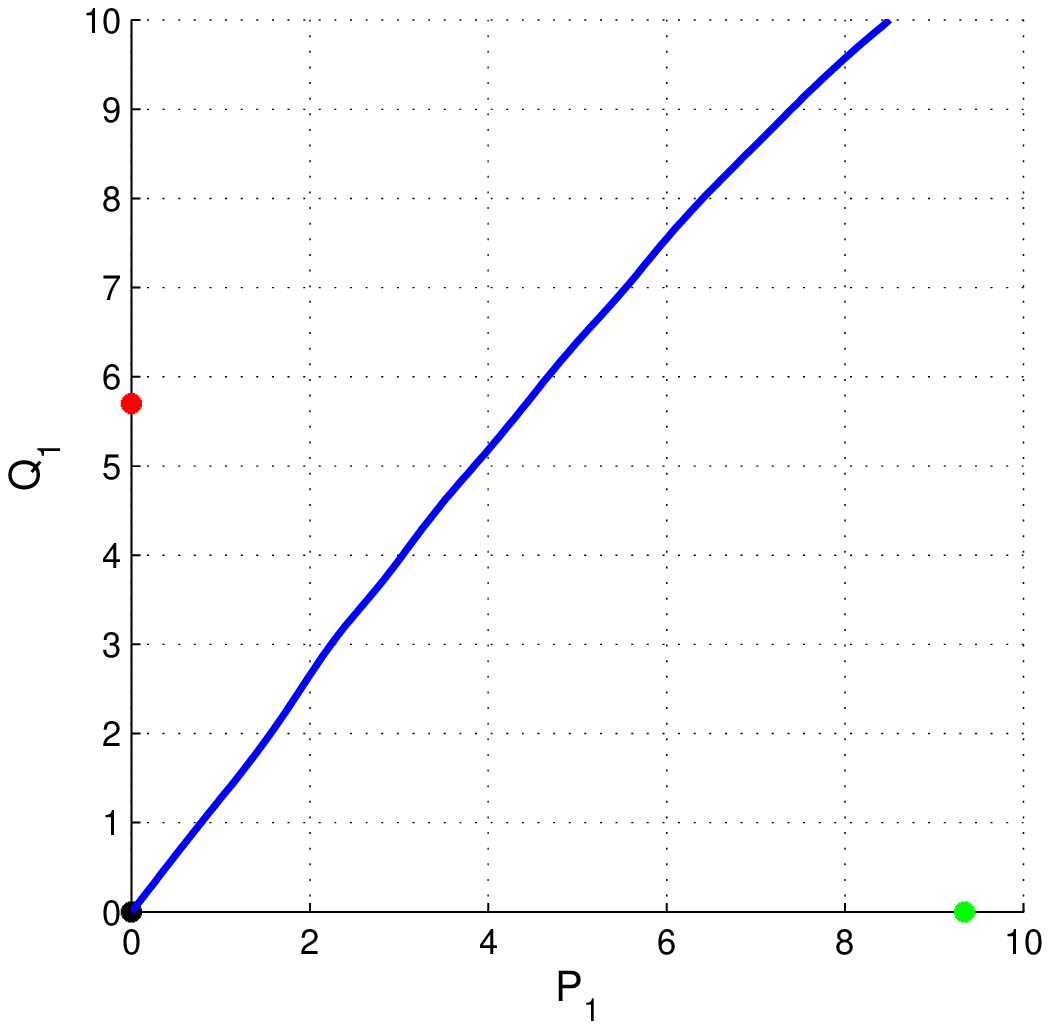}
\includegraphics[width=5.0cm]{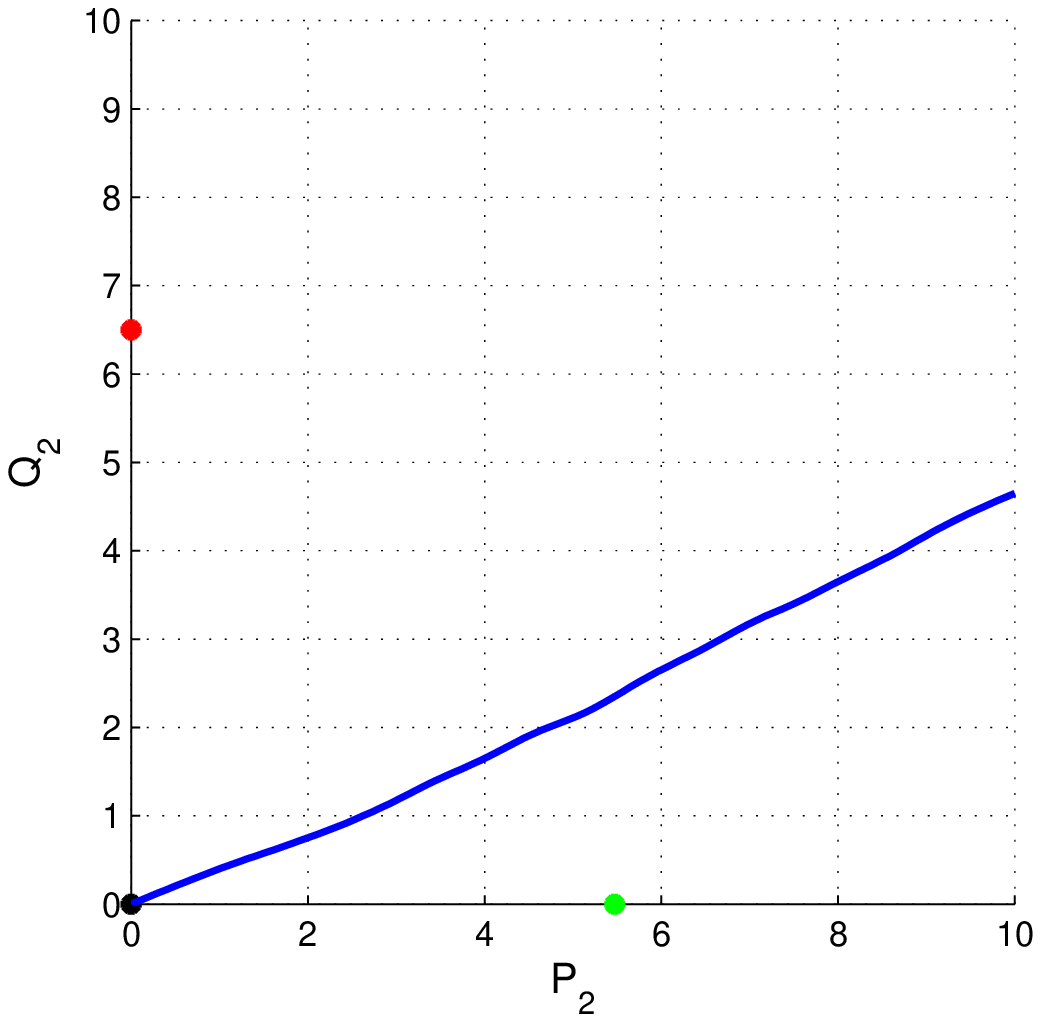}
\caption{Only population $P$ is able to migrate:
separatrix of the basins of attraction of the equilibria $\widehat E_6$ and $\widehat E_{11}$ lying on the axes.
The demographic parameters are given by (\ref{param_demog}).
Right column: patch 1; left column: patch 2.
Top:
% obtained for the following migration rates:
$m_{21} = 0.1$, $m_{12} = 0.1$, $n_{21} = 0$, $n_{12} = 0$.
Bottom:
% obtained from the paramter values
$m_{21} = 0.1$, $m_{12} = 2.0$, $n_{21} = 0$, $n_{12} = 0$.
}
\label{fig:separ_onemigr}
\end{figure}

For this model where just one population is allowed to migrate,
keeping the following demographic parameters fixed,
\begin{eqnarray}\label{param_demog}
r_1 = 6.4, \quad   
s_1 = 5.0, \quad   
k_1 = 8.0, \quad   
h_1 = 5.7, \quad   
a_1 = 2.9, \quad   
b_1 = 2.5, \\ \nonumber
r_2 = 5.5, \quad 
s_2 = 4.6, \quad 
k_2 = 8.2, \quad 
h_2 = 6.5, \quad 
a_2 = 2.3, \quad 
b_2 = 1.7,
\end{eqnarray}
and using the following migration rates
$$
m_{21} = 0.1, \quad   
m_{12} = 0.1, \quad
n_{21} = 0, \quad   
n_{12} = 0,
$$
we have respectively the following stable equilibria
$\widehat E_6 = (8.0057, 0, 8.1962, 0)$, $\widehat E_{11} = (0, 5.7, 0, 6.5)$. The separatrices are
pictured in the top row of Figure \ref{fig:separ_onemigr}, the right frame containing
patch 1 and the left one patch 2.
If we change the migration rates, allowing a faster return toward patch 1,
$$
m_{21} = 0.1, \quad 
m_{12} = 2.0, \quad 
n_{21} = 0, \quad 
n_{12} = 0,
$$
the second equilibrium $\widehat E_{11}$ remains unchanged, but we find instead that the point
$\widehat E_6 = (9.3399, 0, 5.4726, 0)$ has moved toward higher $P_1$ and lower $P_2$ population values.
The separatrices are plotted in the bottom row of Figure \ref{fig:separ_onemigr}.
It is also clear that the basins of attraction in patch 1 hardly change, while in patch 2
the basin of attraction of the population $Q_2$ appears to be larger with a higher emigration rate from
patch 2. Correspondingly, the one of $P_2$ becomes smaller in patch 2, according to what intuition
would indicate.

\subsection{Unidirectional migrations.}
When migrations are allowed from patch 1 into patch 2 only,
a number of other possible equilibria arise, in part replacing some of the former ones. Granted that
coexistence is once again forbidden for its instability, three new equilibria arise, containing either one or both populations in the patch
toward which migrations occur, leaving the other one possibly empty. The principle of competitive exclusion in this case may still occur at the
metapopulation level, but apparently coexistence at equilibrium $\widetilde E_4$ might be possible
in the patch toward which populations migrate
if the stability conditions (\ref{tildeE4_stab}) coupled with the feasibility conditions (\ref{tildeE4_feas_b}) are satisfied.
This appears to be also an interesting result.

\begin{figure}[p]
\centering
\includegraphics[width=5.0cm]{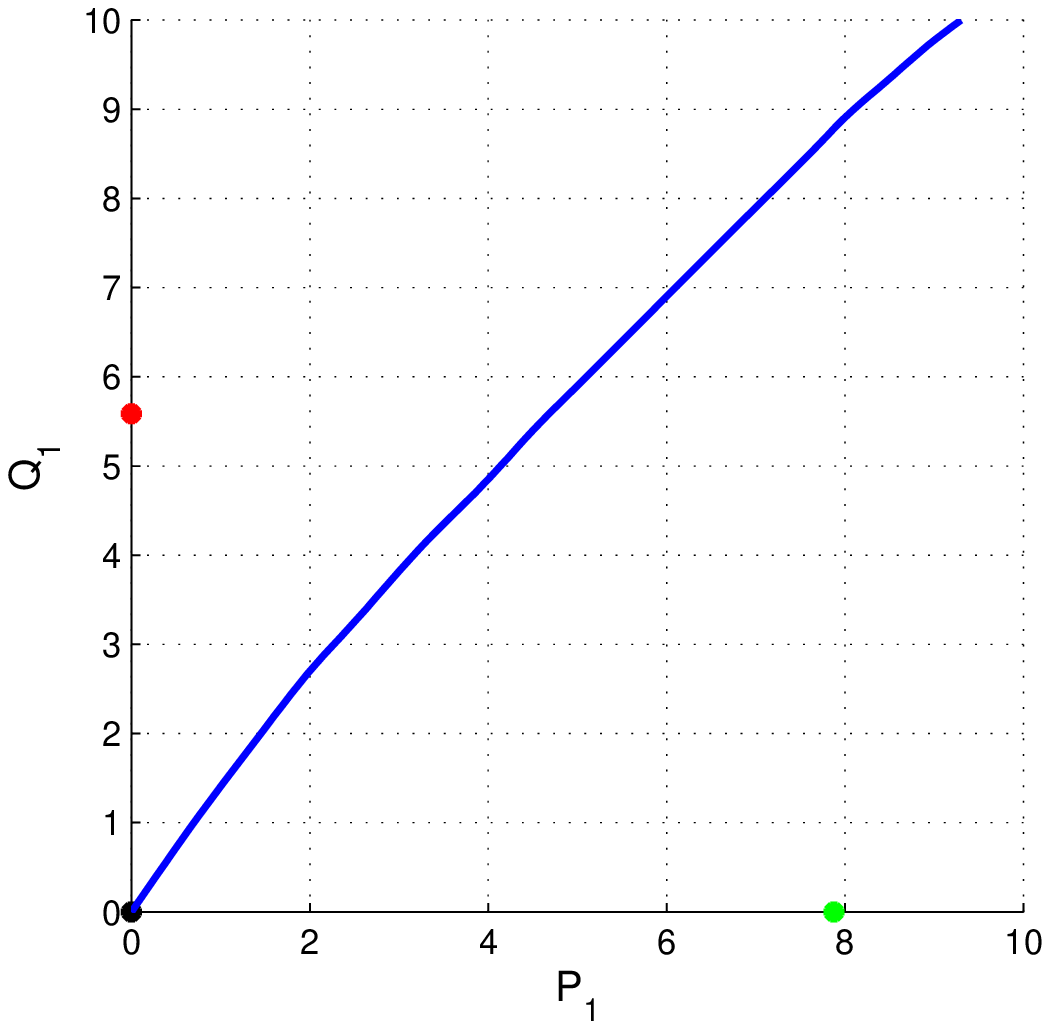}
\includegraphics[width=5.0cm]{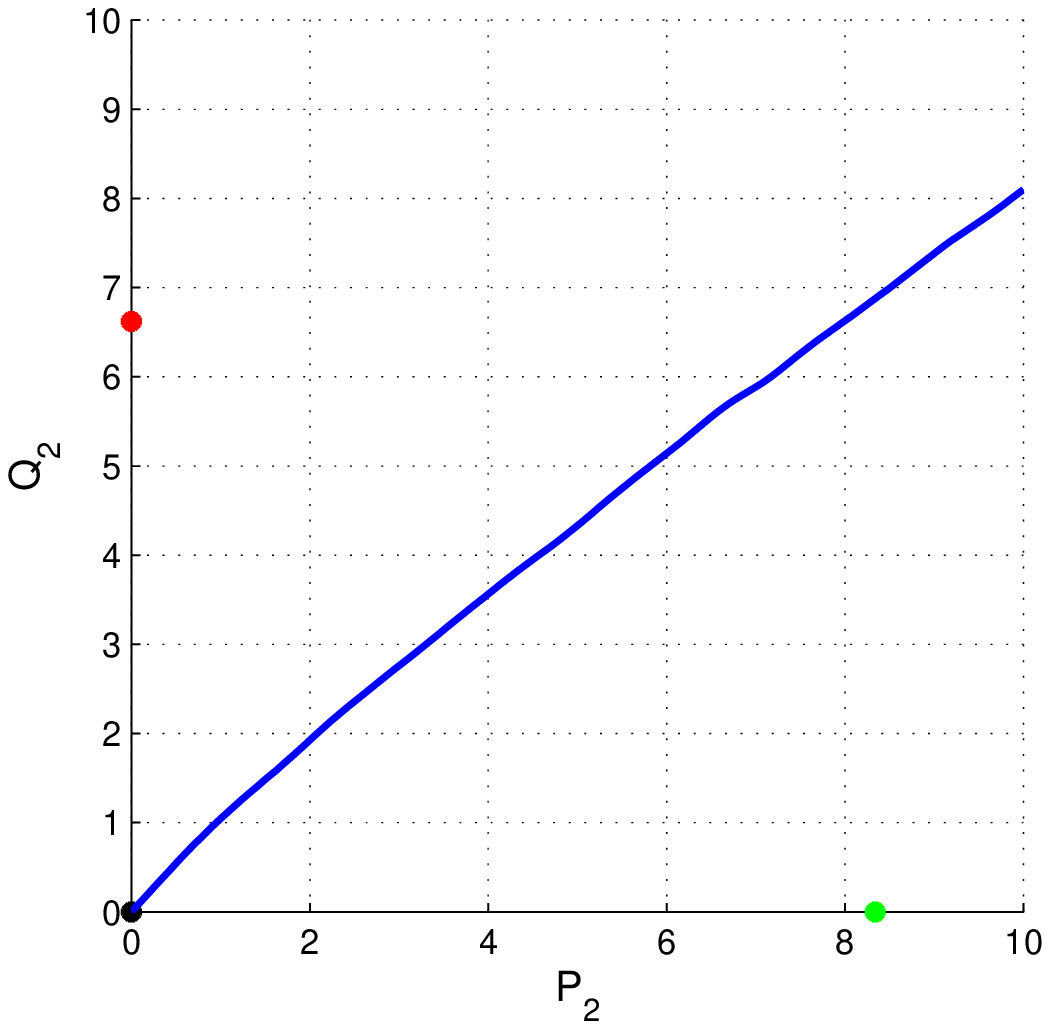}
\includegraphics[width=5.0cm]{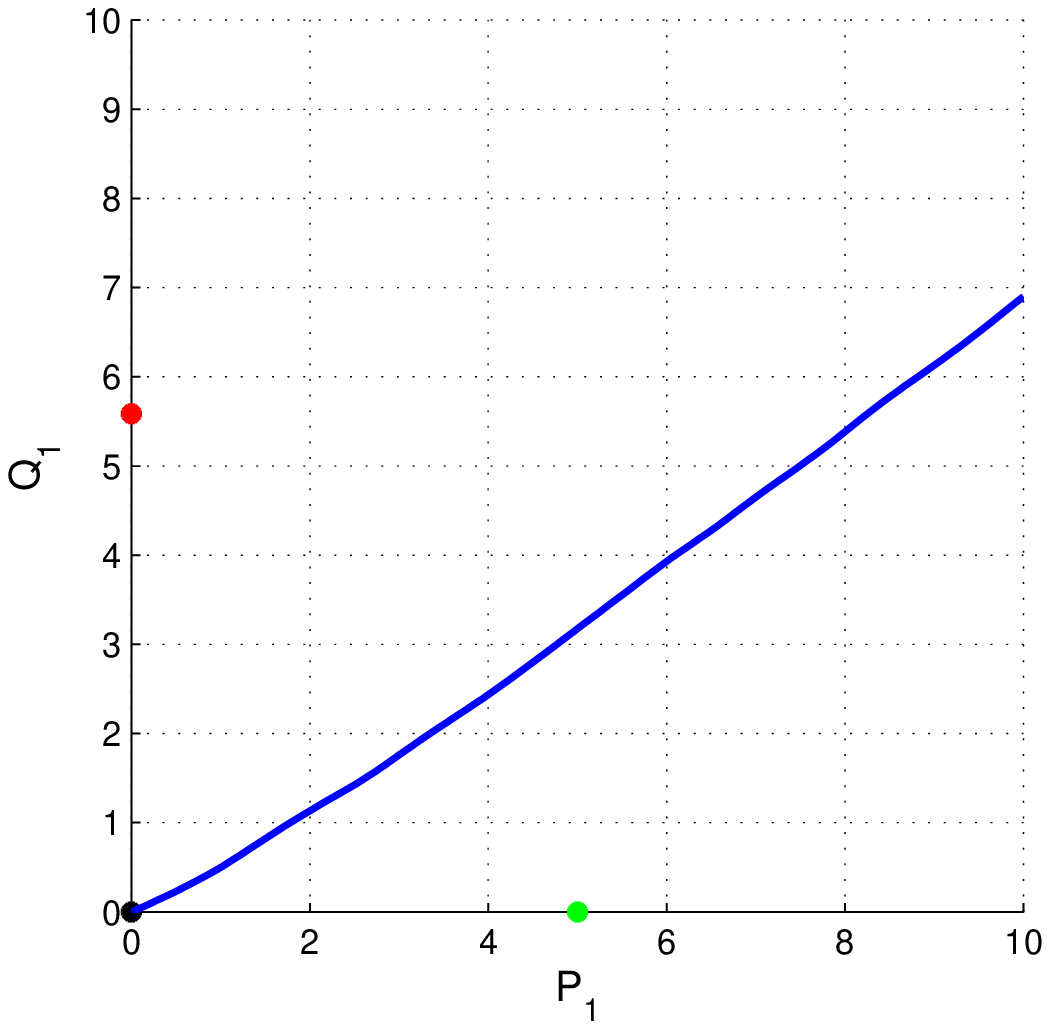}
\includegraphics[width=5.0cm]{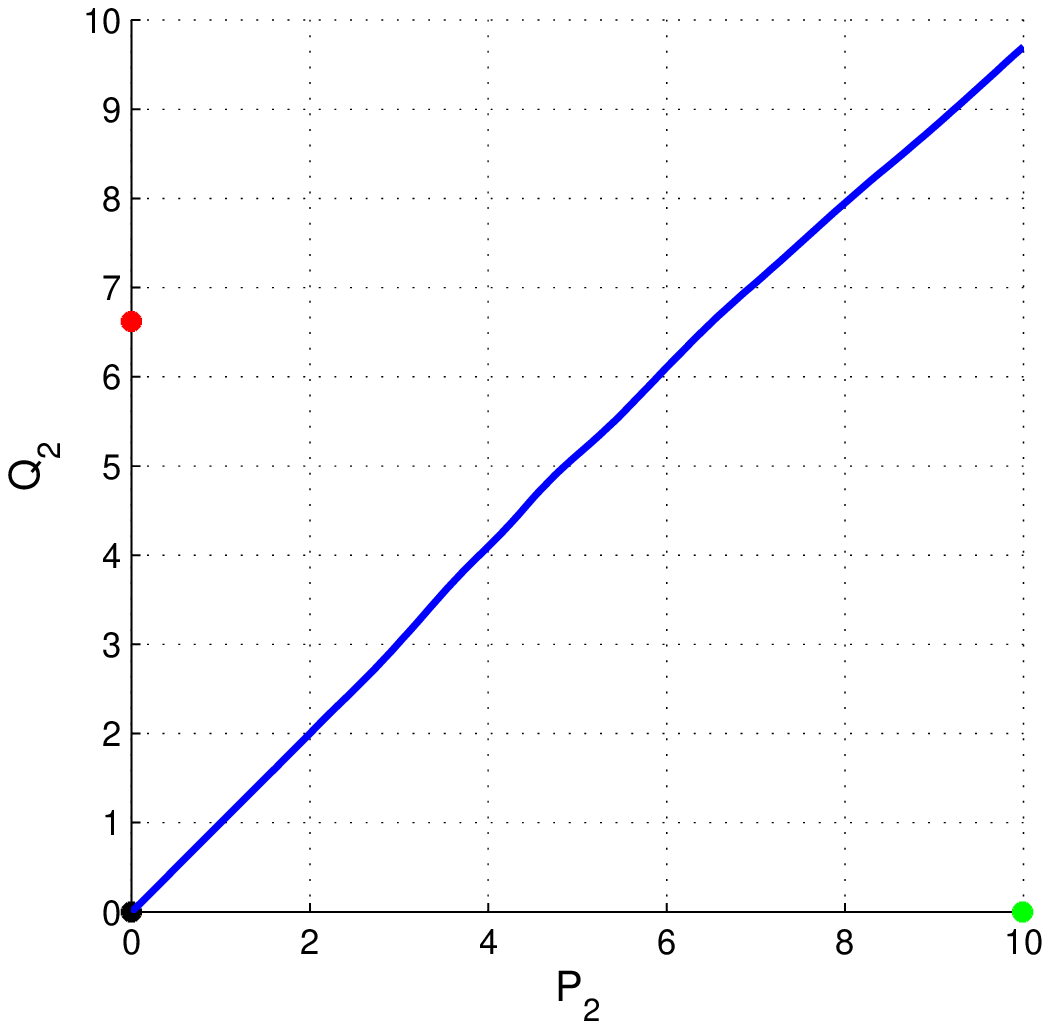}
\caption{Unidirectional migrations:
separatrix of the basins of attraction of the equilibria $\widetilde E_6$ and $\widetilde E_{11}$ lying on the axes.
Demographic parameters are given by (\ref{param_demog}).
Right column: patch 1; left column: patch 2.
Top:
% obtained for the following migration rates:
$m_{21} = 0.1$, $m_{12} = 0$, $n_{21} = 0.1$, $n_{12} = 0$.
Bottom:
% obtained from the paramter values
$m_{21} = 2.4$, $m_{12} = 0$, $n_{21} = 0.1$, $n_{12} = 0$.
}
\label{fig:separ_unidir}
\end{figure}

Again exploiting the algorithm of \cite{CCDRM},
we investigated also the change in shape of the basins of attraction of the two
equilibria $\widetilde E_6$ and $\widetilde E_{11}$,
For this unidirectional migrations model.
Using once again the demographic parameters (\ref{param_demog}),
we take at first the migration rates as follows
$$
m_{21} = 0.1, \quad 
m_{12} = 0, \quad 
n_{21} = 0.1, \quad 
n_{12} = 0,
$$
obtaining equilibria
$\widetilde E_6 = (7.875, 0, 8.3408, 0)$ and $\widetilde E_{11} = (0, 5.5860, 0, 6.6192)$.
This result is shown in the top row of Figure \ref{fig:separ_unidir}, again
patch 1 in the right frame and patch 2 in the left one.
Instead with the choice
$$
m_{21} = 2.4, \quad 
m_{12} = 0, \quad 
n_{21} = 0.1, \quad 
n_{12} = 0,
$$
allowing a faster rate for the population $P$,
we again find that the second equilibrium $\widetilde E_{11}$ is unaffected, but the first one lowers
its population values, becoming
$\widetilde E_6 = (5, 0, 9.9907, 0)$, see bottom row of Figure \ref{fig:separ_unidir}.
In this case the basins of attraction seem to have opposite behaviors. With a higher migration rate for $P_2$,
its basin of attraction in patch 2 gets increased, while in patch 1 becomes smaller. This result is in
agreement with intuition, in patch 1 the $P$ population become smaller and larger instead in patch 2.

\subsection{Final considerations.}
We briefly discuss also the model bifurcations for the unidirectional migration model. 
If $r_1<m_{21}$ and $s_1<n_{21}$, the only feasible equilibria are $\widetilde E_2$, $\widetilde E_3$,
which are stable under the additional conditions $s_2<b_2K_2$ and $r_2<a_2H_2$.
When $r_1$ crosses the value $m_{21}$ and similarly $s_1\ge n_{21}$, the two previous equilibria become unstable,
and transcritical bifurcations give rise
respectively to the equilibria $\widetilde E_6$ and $\widetilde E_{11}$.
The equilibrium $\widetilde E_4$ may coexist with each one
of the previous equilibria, but in this case $\widetilde E_2$
and $\widetilde E_3$ must be unstable, whereas $\widetilde E_6$ and $\widetilde E_{11}$ may be stable if their stability conditions hold.

In the two particular cases above discussed, of just one population allowed to migrate and of
unidirectional migrations,
our analysis shows that the standard assumptions used to study configurations in patchy environments
may not always hold. Under suitable conditions, competing populations may coexist if
only one migrates freely, or if migrations for both populations are allowed in the same direction and not backwards.
This appears to be an interesting result, which might open up new research directions.

\end{document}